 \newlength\smallfigwidth
\documentclass[aps,prb, 10pt,twocolumn,floatfix,showpacs,amsmath,
amssymb,superscriptaddress] {revtex4-1}

\usepackage{amsmath}    
\usepackage{graphicx}   
\usepackage{verbatim}   
\usepackage{color}      
\usepackage{subfigure}  
\usepackage{hyperref}   

\begin{document}

\preprint{UFES}

\title{Emergence of skyrmion lattices and bimerons in diluted chiral magnetic thin films}

\author{R.\ L.\  Silva}
\email{ricardo.l.silva@ufes.br} 
\affiliation{ Departamento de Ci\^encias Naturais, Universidade Federal do
Esp\'irito Santo, S\~ao Mateus, 29932-540, Esp\'irito Santo, Brazil}

\author{L.\ D.\ Secchin }
\email{leonardo.secchin@ufes.br} 
\affiliation{Departamento de Matem\'atica Aplicada, Universidade Federal do
Esp\'irito Santo, S\~ao Mateus, 29932-540, Esp\'irito Santo, Brazil}

\author{W.\ A.\  Moura-Melo}
\email{wamoura.melo@gmail.com} 
\author{A.\ R.\  Pereira}
\email{apereira@ufv.br} 
\affiliation{Departamento de F\'isica, Universidade Federal de Vi\c cosa, Vi\c
cosa, 36570-000, Minas Gerais, Brazil}

\author{R.\ L.\  Stamps}
\email{Robert.Stamps@glasgow.ac.uk} 
\affiliation{SUPA School of Physics and Astronomy, University of Glasgow,
Glasgow G12 8QQ, United Kingdom}

\date{\today}
\begin{abstract}
 Skyrmions are topologically protected field configurations with particle-like properties that play important roles in various fields of science. Recently, skyrmions have been directly observed in chiral magnets. Here, we investigate the effects of nonmagnetic impurities (structural point-like defects) on the different initial states (random or helical states) and on the formation of the skyrmion crystal in a discrete lattice. By using first-principle calculations and Monte Carlo techniques, we have shown that even a small percentage of spin vacancies present in the chiral magnetic thin film affects considerably the skyrmion order. The main effects of impurities are somewhat similar to thermal effects. The presence of these spin vacancies also induces the formation of compact merons (bimerons) in both the helical and skyrmion states. We have also investigated the effects of adjacent impurities producing only one hole (with an almost circular shape) intentionally inserted into the plate (forming a non-simply 
connected manifold) on the skyrmion crystal.
\end{abstract}

\maketitle
\section{Introduction and motivation}
Topological excitations are exciting objects in the physical world. In condensed matter
physics, they frequently arise in the form of vortices and skyrmions which carry,
in general, integer topological charges. However, topological excitations
(more controversial) carrying only one-half of the skyrmion number, the so-called merons, may also be present in condensed matter
systems. In this work, we pay attention to some
topologically stable skyrmion and compact merons (bimerons) spin textures in diluted helical magnets. It is shown that some percentage of impurities introduced in these systems is able to induce the incidence of bimeron excitations and a deformation of the skyrmion lattice. The six fold order is lost for a very low concentration of spin vacancies.

Skyrmions were first found by Skyrme \cite{Skyrme61} about $50$ years ago during an
investigation of classical Lagrangian for possible models of subatomic particles as
neutrons and protons (nucleons as discrete entities emerging from a continuous nuclear field).
Despite skyrmion was originally introduced in field theory, it is now highly
relevant to spin structures in condensed-matter systems as well. For
example, such ideas were introduced in the context of the two-dimensional ($2D$)
magnetism by Belavin and Polyakov \cite{BP75} ($BP$). These authors considered the continuum limit of the isotropic Heisenberg magnet, i.e., the nonlinear $\sigma$ model ($NL\sigma M$), which is described by the Hamiltonian $H_{\sigma}=J/2 \int [(\partial_{\alpha}\vec{\mu})\cdot \partial_{\alpha}\vec{\mu})]d^{2}x$, where $J$ is the ferromagnetic coupling
constant and $\mu=\mid \vec{\mu} \mid $ is the spin of the system (obeying
the constraint $\vec{\mu}^{2}=\mu_{x}^{2}+ \mu_{y}^{2}+ \mu_{z}^{2}=1$ ). In this framework, skyrmion-like objects show up as topological solitonic excitations carrying finite energy. Topologically, they correspond to the mapping of the spin-space sphere
$\sum^{2}$ onto the continuum plane $\Re^{2}$. Their energy reads $E_{s}=4\pi QJ\mu^{2}$, where $Q=\int (d^{2}r/8\pi) \epsilon_{ij} \epsilon_{\alpha \beta \gamma}\mu_{\alpha}\partial_{i}\mu_{\beta}\partial_{j}\mu_{\gamma}$
is the topological charge, counting the number of times the
internal sphere $\sum^{2}$ is traversed as one span the physical space
$\vec{r}=(x,y)$ (as compacted in a sphere). For $Q=1$, a $BP$ skyrmion of
size $R$ has the well known configuration:
\begin{eqnarray}\label{Skyrmion1}
\mu_{z}=\frac{r^{2}-R^{2}}{r^{2}+R^{2}}       ,\,\,\,\,\,\,\,
\mu_{x}+ i \mu_{y}=2iR\frac{x+iy}{r^{2}+R^{2}}
\end{eqnarray}
Note that the skyrmion energy does not depend on the skyrmion size $R$, which is a signature of the scale invariance of the $NL\sigma M$.

Unfortunately, evidences for $BP$ skyrmions in layered isotropic classical magnets (not chiral) are only indirect. To our knowledge, the first indirect observation of such excitations in quasi-$2D$ nearly classical magnets (a material with spin-$5/2$, described by the Heisenberg Hamiltonian) was reported by Waldner\cite{Waldner86,Waldner90} who found experimentally the skyrmion energy ($E_{exp}$ ) from the heat capacity measurements in good agreement with theoretical prediction by $BP$ ($E_{exp}\simeq E_{s}$ ). Further indirect observations of skyrmions were reported in quasi-$2D$ manganese compounds (systems with spin-$5/2$) and other class of isotropic classical antiferromagnets (propylammonium tetrachloromanganate also with spin-$5/2$), in which the skyrmion traces were found in the Electron Paramagnetic Resonance ($EPR$) linewidth (see Zaspel \textit{et al.}\cite{Zaspel95}). Such calculations were based on the skyrmion-magnon interactions which also resulted in an experimental skyrmion energy $E_{exp}\
simeq E_{s}$. All these works (Waldner using a thermodynamic quantity and Zaspel \textit{et al.} using a dynamic quantity) considered only indirect evidences for isolated skyrmions as excited states (in general, the isotropic Heisenberg magnets are expected to contain a diluted ideal gas of these excitations).

Recently, the discovery and real-space observations of the skyrmion phase of magnetization in magnetic materials have renewed the interest in these objects in condensed matter physics, calling great attention of the physics community. Indeed, important direct experimental works\cite{Yu10,Heinze11} have undoubtedly revealed, for the first time, the very presence of skyrmions in chiral magnets, where the Dzyaloshinsky-Moriya ($DM$) interaction plays a fundamental role (chiral materials are systems in which the mirror reflection symmetry is broken). However, differently from the above cases, in which isolated skyrmions (in a diluted gas of skyrmions) were indirectly considered in isotropic magnets, in the chiral magnets, they can be seen either in regular lattices or even as topologically stable single particles. Besides of being directly observed  at low temperatures in films of $Fe_{0.5}Co_{0.5}Si$ they also appear close to room temperature in $FeGe$. Actually, these compounds share the same $B20$ structure 
with chiral cubic symmetry and they commonly show the helimagnetic ground state. Whenever a magnetic field is applied perpendicularly to the sample plane a $2D$ skyrmion lattice emerges \cite{Yu10}. On the other hand, in iron films (with $DM$ interaction), skyrmions form spontaneously as the ground state, without necessity of being stabilized by any small external magnetic field \cite{Heinze11}. Although the exact spin configuration of a single skyrmion in the skyrmion-lattice is somewhat different from that given in Eq.(\ref{Skyrmion1}), the topological properties must remain basically the same, rendering the robustness of skyrmion-lattice as experimentally observed. Moreover, in a lattice, the skyrmion size $R_{L}$ is uniquely determined by the ratio of $DM$ and exchange energies (similarly, for skyrmions in the quantum Hall systems, the skyrmion size is fixed by the relative strengths of Zeeman and Coulomb interactions).

\begin{figure}[hbt]
\begin{center}
\subfigure[]{
\includegraphics[height=41.0mm,width=41.0mm]{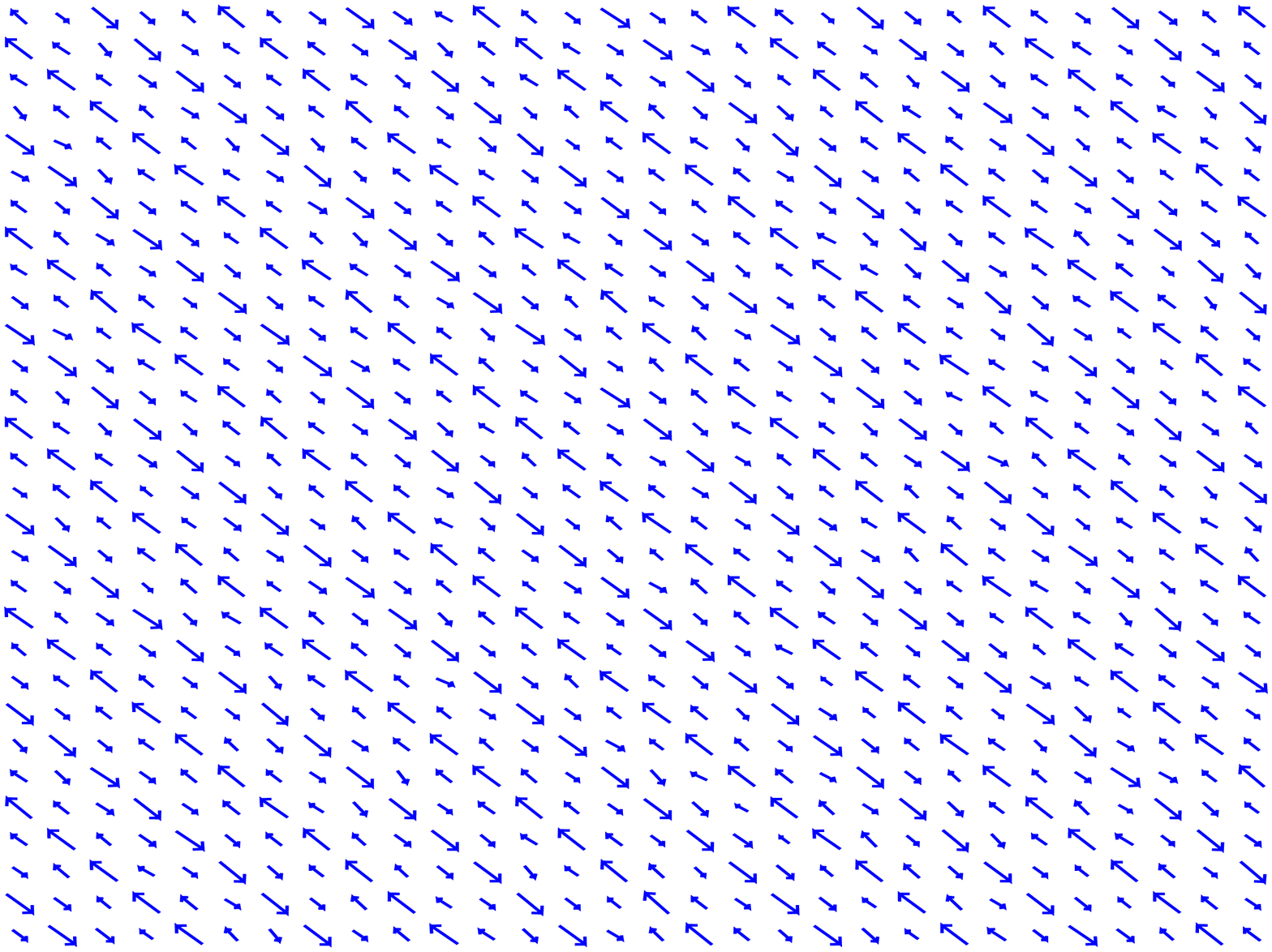}}
\subfigure[]{
\includegraphics[height=41.0mm,width=41.0mm]{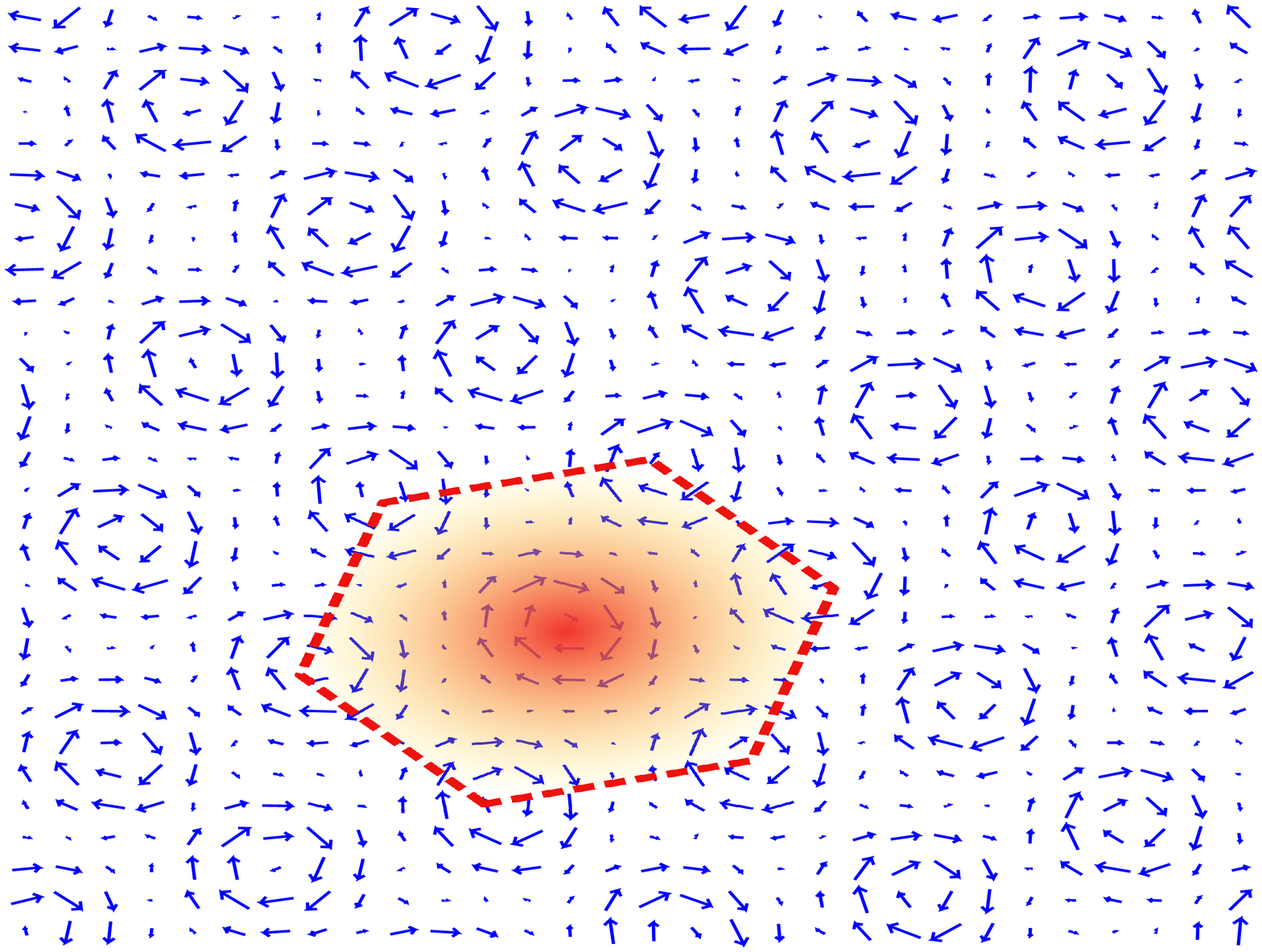}}
\subfigure[]{
\includegraphics[height=41.0mm,width=41.0mm]{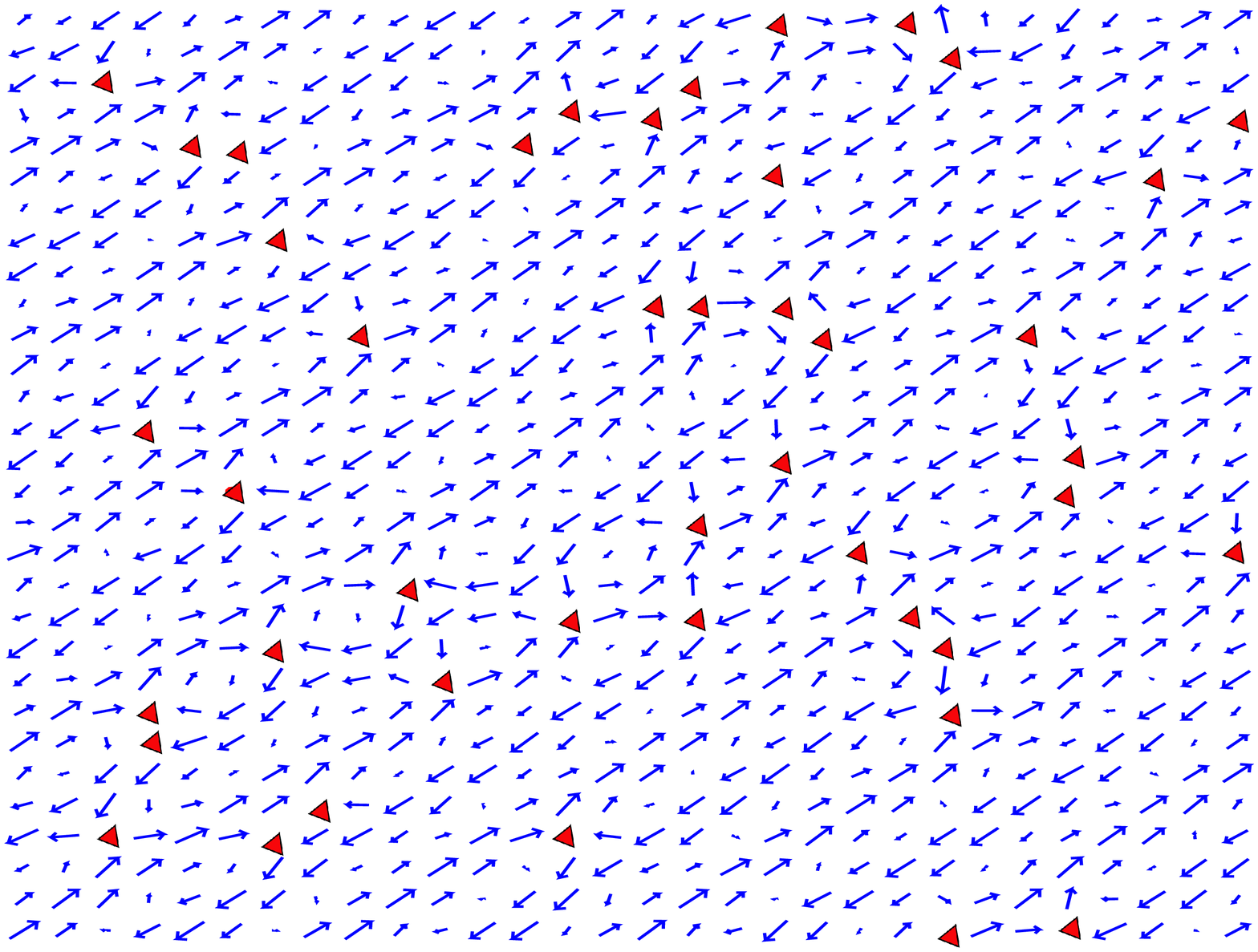}}
\subfigure[]{
\includegraphics[height=41.0mm,width=41.0mm]{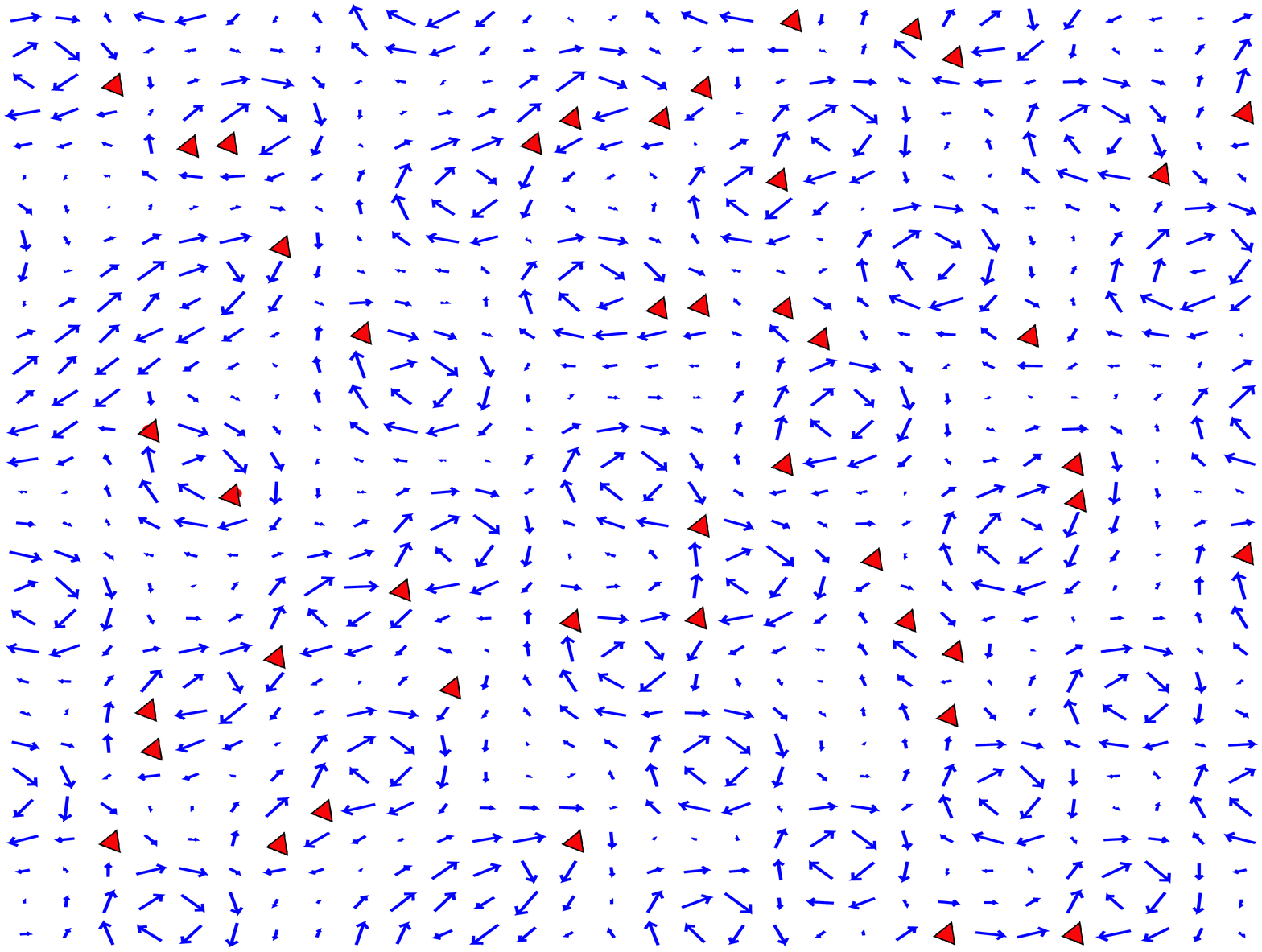}}
\caption{\label{fig:qm/l1}  (a): Spiral State in pure systems. (b): Hexagonal
Skyrmion Crystal submitted to a a magnetic field $h_{z}$ and free of
impurities. (c): Spiral State with $5$ percent of impurities.
(d): Hexagonal Skyrmion Crystal submitted to a magnetic field $h_{z}$ in a system with
$5$ percent of impurities.}
\end{center}
\end{figure}

On the other hand, a meron-type excitation contrasts with typical topological defects (such as vortices) in easy-plane magnets, once it carries half skyrmion number $\pm 1/2$ and vanishing vorticity. Thus, a pair of merons (a bimeron) is rather similar to a skyrmion, carrying integer skyrmion charge. The existence of merons in helical magnets is still less evident. Compact merons (or bimerons) have been proposed in the context of chiral magnets by Ezawa \cite{Ezawa11}. These objects are expected to arise above the ground state of these systems (the stripe helical state) whenever a stripe breaks into pieces as the magnetic field increases; the end points of a broken stripe bears opposite half skyrmion charges ($\pm 1/2$) but the vortex number is zero. Thus, a finite stripe becomes a a pair of merons. The bimeron is then composed by two half-disk domains and a rectangular stripe domain connecting these half-disks \cite{Ezawa11}. The bimeron is then composed by two half-disk domains and a rectangular stripe 
domain connecting these disks \cite{Ezawa11}. Consequently, the spin texture of a compact skyrmion could be obtained simply by removing the rectangular part of a bimeron and by joining the two half-disk domains. Ezawa \cite{Ezawa11} has demonstrated that a treatment of merons as free particles qualitatively captures the phase diagram as predicted by Monte Carlo simulation \cite{Ezawa10} and observed in experiments \cite{Yu10}. Our results not only reinforce the existence of these compact merons in chiral magnets, but they also indicate that such objects should appear pinned by non-magnetic impurities throughout the whole sample, even at the absence of external field.

In this work we show that, in the presence of an impurity density $\rho$ (around a few percents), the stripe helical state undergoes a transition to a state where finite stripes show up through the sample; additionally, the skyrmion lattice state is also drastically affected, namely, its hexagonal-like collective arrangement appears to be deeply changed even at very low $\rho$. Figures \ref{fig:qm/l1}$(a)$ and \ref{fig:qm/l1}$(b)$ illustrate the ground state and the skyrmion crystal (respectively) for pure systems while Figs. \ref{fig:qm/l1}$(c)$  and \ref{fig:qm/l1}$(d)$ show how these same states change in a film with $5\%$ of impurities. In more general grounds, it is known that a spin vacancy in $2D$ nearly classical isotropic magnets generates an interaction potential that attracts the skyrmion center \cite{Subbaraman98,Mol03,Pereira03}. Hence, in these magnets containing a low impurity density, such potential allows a vacancy to be at the skyrmion center (similarly, these kind of defects also attract 
the vortex center in easy-plane magnets\cite{Pereira05} and magnetic nanodisks \cite{Rahm03,Paula04,Silva08,Pereira07}). Particularly, in easy-plane magnets, a low concentration of nonmagnetic impurities has a strong influence on the Berezinskii-Kosterlitz-Thouless ($BKT$) transition temperature \cite{Wysin05} and in magnetic nanostructures they generate the physical mechanism behinds the experimental observation of the Barkhauser effect \cite{FelipeJAP2009,Science2013-exp}. Extrapolating these ideas, we could argue that possible interactions between these topological objects and lattice defects (such as spin vacancies) may distort the helical state and the skyrmion lattice, changing the skyrmion center positions in a nontrivial way. We then investigate the properties of the skyrmion crystal and the percentages of impurities necessary for completely destroying the six fold order and the skyrmion itself. For the helical state, we observe that vacancies stimulate the creation of bimerons above the perfect 
stripe-domain structure. Vacancies may then develop a role rather similar to thermal effects \cite{Ambrose12} and also comparable to magnetic field cooling \cite{Science340}, causing the creation of excitations with elongated linelike patterns \cite{Ambrose12,Science340} and a kind of a skyrmion melting, even at zero temperature, for a given critical concentration of lattice defects.

\section{Methods and Results}
In order to describe our system, we explore the magnetic phase space on the basis of a minimal lattice Hamiltonian, which is the following extended two-dimensional Heisenberg model \cite{Ambrose12}:
\begin{eqnarray}\label{eq:h1}
H &=&-J\sum_{\vec{r}}\vec{\mu}_{\vec{r}}\cdot (\vec{\mu}_{\vec{r}+\hat{x}}+\vec{\mu}_{\vec{r}+\hat{y}})-D\sum_{\vec{r}} (\vec\mu_{\vec{r}}\times\vec\mu_{\vec{r}+\hat{x}}\cdot\hat{x}\nonumber\\
&& +\vec\mu_{\vec{r}}\times\vec\mu_{\vec{r}+\hat{y}}\cdot \hat{y}) +A_{1}\sum_{i}((\mu_{\vec{r}}^{x})^{4}+(\mu_{\vec{r}}^{y})^{4}+(\mu_{\vec{r}}^{z})^{4})\nonumber\\
&& +A_{2}\sum_{i}(\mu_{\vec{r}}^{x}\mu_{\vec{r}+\hat{x}}^{x}+\mu_{\vec{r}}^{y}\mu_{\vec{r}+\hat{y}}^{y})  -\sum_{\vec{r}}\vec{h}_{\vec{r}}\cdot\vec{\mu}_{\vec{r}}\textnormal{.}
\end{eqnarray}
Here $\vec{\mu}_{\vec{r}}=\mu_{\vec{r}}^{x}\hat{x}+\mu_{\vec{r}}^{y}\hat{y} + \mu_{\vec{r} }^{z}\hat{z}$ is the unit spin vector at position $\vec{r}$ and the first sum is over nearest-neighbor spins with exchange interaction strength $J>0$. The second term is the Dzyaloshinskii-Moriya interaction, where $\vec\mu_{\vec{r}+\hat{x}}$ and $\vec\mu_{\vec{r}+\hat{y}}$ represent the closest spins to $\vec{\mu}_{\vec{r}}$ in the directions $x$ an $y$ respectively. The third and fourth terms, with $A_{1}$ and $A_{2}$ constants, consist in the anisotropy of the system. The last term considers the effects of an external magnetic field $\vec{h}$ (Zeeman energy).

\begin{figure}[hbt]
\begin{center}
\includegraphics[width=84mm]{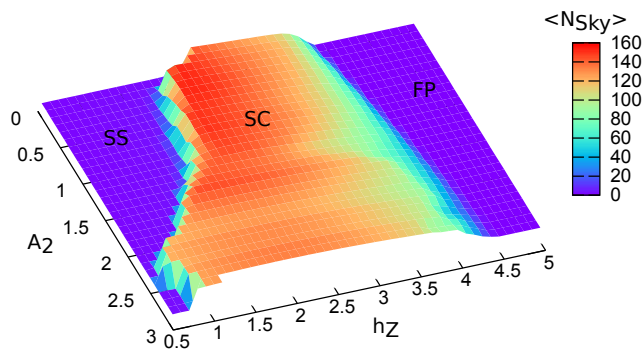}
\caption{\label{fig:qm/l2} Plot of the Phase diagram:
Spiral State (\textbf{SS}), Skyrmion Crystal (\textbf{SC}) and Field-
Polarized (\textbf{FP}) states.}
\end{center}
\end{figure}

Firstly, we consider the ground states obtained from Eq. (\ref{eq:h1}). To do
this, we use a simulated annealing process, which is a Monte Carlo
calculation where the temperature is slightly reduced in each step of
the process in order to drive the system to the global minimum. Our
Monte Carlo scheme consists of a simple Metropolis algorithm. In each
Monte Carlo step ($MCS$), we attempt to flip all spins in the lattice
sequentially or randomly, which gives the same results. From an initial
spin configuration obtained by randomly occupying sites with probability $1 -
\rho$, the simulation is performed with periodic boundary conditions for system
size $N = L^2$; here, we have investigated lattice sizes with $L = 40$, $60$
and $80$. Then, the number of magnetically occupied sites is $N_{mag}
= N (1 - \rho)$. In each simulation $10^{5}\times L^2$ MCSs were done
at each temperature, starting at $T=1.0$; the temperature is decreased
in steps $\delta T=0.01$ until $T=0.01$ (the temperature is measured
in units of $J/k_{B}$). We choose the ratio $D/J=\sqrt{6}$ (that
determines the hexagonal periodicity of the skyrmion lattice) and
$A_{1}=0.5$ (without loss of generality). Two techniques have been employed
to measure the number of skyrmions in a state \cite{Ambrose12}: the first one takes into account
the topological charge $Q$ of a skyrmion, which is the sum of the local chiral charge over
the area of a single skyrmion, $Q_{sky} = \int{d^{2}x\rho_{sky}}$, with
\begin{equation}
\rho_{sky}=\frac{1}{4\pi}\vec{\mu}_{\vec{r}}\cdot
((\vec{\mu}_{\vec{r}+\hat{x}}-\vec{\mu}_{\vec{r}-\hat{x}})\times
(\vec{\mu}_{\vec{r}+\hat{y}}-\vec{\mu}_{\vec{r}-\hat{y}}))
\end{equation}
where $\rho_{sky}$ is the topological charge density for several spin
textures. In order to define the charge $Q$ of a skyrmion, one
calculates the total charge density of the ground state and divides by the
number of skyrmions. To calculate the number of skyrmions at
some finite temperature one divides the total charge by $Q$. The
second method \cite{Ambrose12} consists in applying $\Theta(1/2(\vec{\mu}_{\vec{r}}\cdot
\hat{z}+1)-\Gamma)$ ($\Theta$ is the Heaviside Theta function), were
$\Gamma$ is some threshold between zero and one. Since the core of a
skyrmion points against the applied magnetic field, we can binarize a
state by applying the function $\Theta$ identifying spins with
$\vec{\mu}_{\vec{r}}\cdot\hat{z}$ in the core of Skyrmions. By
comparing these two measurements of Skyrmion number one can distinguish
between reduction of topological charge due to Skyrmion destruction
and alteration in of Skyrmion profile (for more details, see Ref. \cite{Ambrose12}).

\begin{figure}[hbt]
\begin{center}
\subfigure[]{
\includegraphics[height=80.0mm]{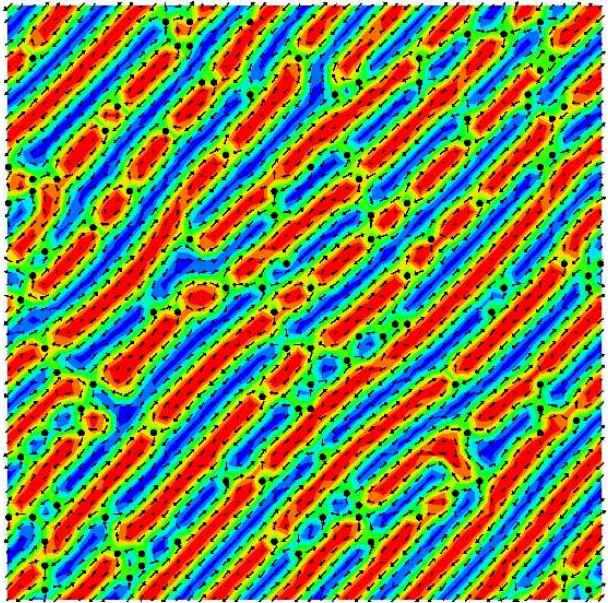}}
\subfigure[]{
\includegraphics[height=80.0mm]{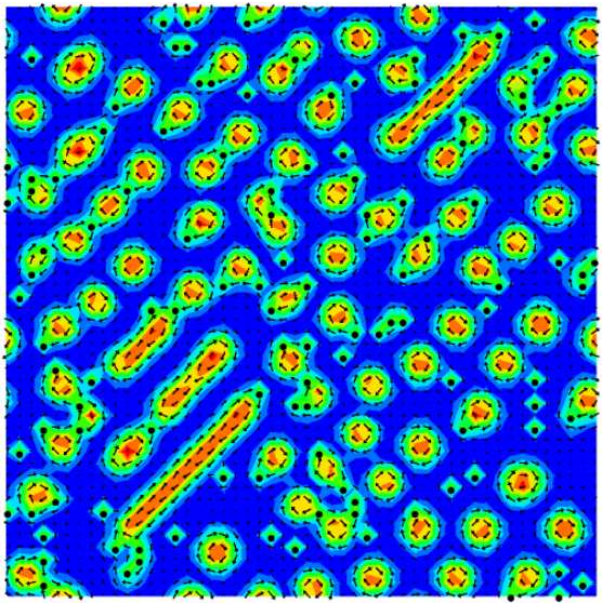}}
\caption{\label{fig:qm/l5}  (a): Spiral State with $5\%$ of
impurities. (b): Hexagonal Skyrmion Crystal submitted to a magnetic
field $h_{z}$ in a system with $5\%$ of impurities. The
background color represent the out-of-plane component of the local
magnetic moment ($\vec{\mu}_z$) with blue for $\mu_z=-1$, red for $\mu_z=+1$
and green for $\mu_z=0$. The black dots indicate the impurities
positions.}
\end{center}
\end{figure}

Before inserting impurities, we have obtained the phase diagram (for pure systems) by
decreasing the temperature (in zero field) to achieve the stripe helical
state structure for each $A_{2}$ (varying in the interval $0\leq
A_{2} \leq 2.0$). When the perpendicular external field
$\vec{h}=h_{z}\hat{z}$ is applied, we vary the field in the interval ($0\leq h_{z} \leq
5.0$) with steps $\delta h_{z}=0.01$. The phase diagram spanned by ($A_{2}, h_{z}$) is shown in Fig. \ref{fig:qm/l2}.

\begin{figure}[hbt]
\begin{center}
\includegraphics[width=86mm]{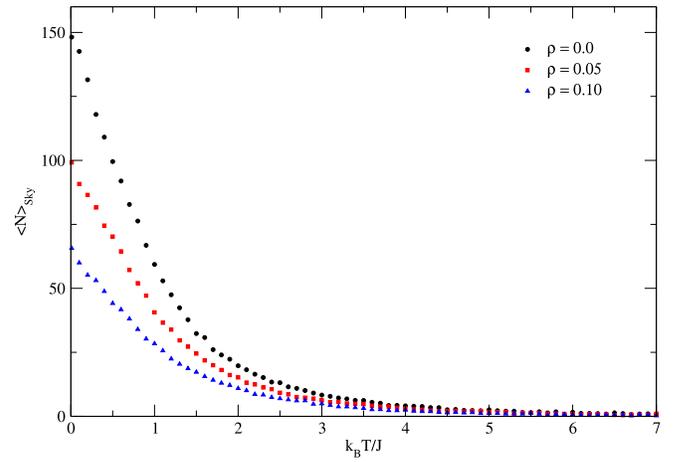}
\caption{\label{fig:qm/l3} Plot of the Skyrmions number $<N>_{sky}$
with varying $T$ for $\rho=0$, $\rho=0.05$ and $\rho=0.1$. }
\end{center}
\end{figure}

\begin{figure}[hbt]
\begin{center}
\includegraphics[width=86mm]{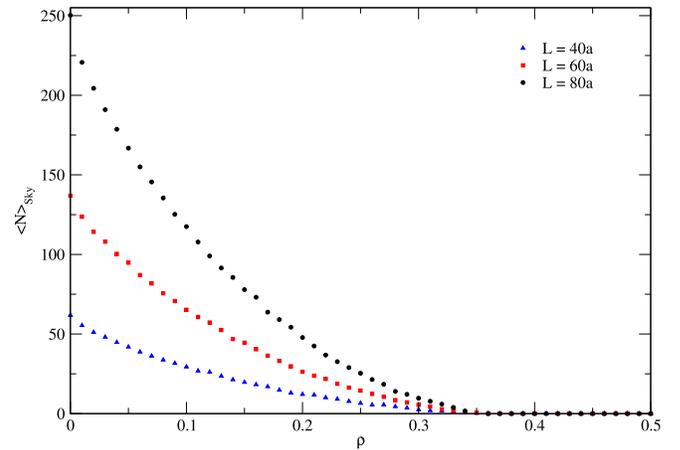}
\caption{\label{fig:qm/l4} Plot of the Skyrmions number $<N>_{sky}$
with varying impurities strength $\rho$ for $L=40a$, $60a$ and
$80a$, at temperature $T=0.01$.}
\end{center}
\end{figure}

To start our investigation, we chose $A_{2}=0.5$ and $h_z=2.0$ (which are the lowest values that lead to skyrmions crystal ($SC$) with hexagonal symmetry ) and add impurities (randomly) into the system. We
find the stripe helical state with impurities and calculate both the number of skyrmions as a function of temperature (with and without impurities) and the number of skyrmions as a function of impurities density.

Firstly, we discuss the stripe helical state. It is interesting to note
that impurities are able to create bimerons in
this state. As shown in Fig.\ref{fig:qm/l1}$(c)$ (for a system with $5\%$ of
impurities), bimerons are the elongated  configurations that look like a
skyrmion but with its central core (which has usually the shape of a disk)
shared in two half-disks separated by a rectangular stripe domain. To better see
such structures, Fig. \ref{fig:qm/l5}$(a)$ presents the helical phase in colors:
blue for spins in the negative $z$-direction ($S_{z}=-1$), green for spins point
out along the $xy$-plane ($S_{z}=0$) and red implies $S_{z}=1$. Therefore, the
helical state deformation comes about due to the presence of bimerons, which are
localized between pieces of a stripe containing spin vacancies. Actually, an
appreciable number of broken stripes show up even for very low impurity
concentration ($\sim 1\%$), rendering impurities as an intrinsic mechanism to
favor the appearance of bimerons in chiral magnets. Besides, impurities also
offer an alternative experimental route towards their observation and even their
precise localization, namely, if usual techniques, as magnetic force microscopy
(MFM), is combined with magneto-mechanical setups, as those used to probe
Barkhausen effect due to spin vancancies in magnetic nanostrucutures
\cite{Science2013-exp}. In this sense such vacancies play a role similar to
magnetic field (as suggested by Ezawa \cite{Ezawa11} for pure samples, and
observed in field cooling processes by Milde {\it et al} \cite{Science340}) and
also to thermal effects, as found by Ambrose and Stamps \cite{Ambrose12}.
Furthermore, these meron pairs will not be organized in regular lattices; they
should be located around impurities randomly distributed throughout the sample.
Additionally, in contrast to the behavior of the usual excitations in the $2D$
isotropic magnets, the bimerons obtained in our simulation do not have their
half-disks centered at spin vacancies; they rather localize along the diagonal
direction of the $xy$-plane and they are, in general, located between two spin
vacancies (in the same stripe, namely when these vacancies are separated by
relatively large distances).

Now we consider the skyrmion crystal state when nonmagnetic impurities are present (see Figs. \ref{fig:qm/l1}$(c)$ and  \ref{fig:qm/l5}$(b)$). Figure \ref{fig:qm/l3} shows the dependence of the skyrmion
number $<N>_{sky}$ as a function of temperature (considering three values of impurity concentration $\rho=0$, $0.05$ and $0.1$). Ambrose and Stamps \cite{Ambrose12}) have studied such thermal behavior of the
skyrmion lattice (and the resulting skyrmion melting) for the case in
which the system is pure ($\rho=0$). These authors have found that the
thermal distortion of the skyrmion profiles (rather than destruction
of the skyrmions themselves) starts to occur at relatively low temperature ($T \approx 0.3
J/k_{B}$). At higher temperatures, there is a sharp loss of the six
fold order associated with the creation of elongated structures that
disrupt ordering. In what follows, we argue that the effects due to the presence of vacancies
in the system lattice is somewhat similar to that of thermal effects as found in Ref.~\onlinecite{Ambrose12}.
Indeed, it distorts the skyrmion profiles, decreasing the skyrmion number as the impurity
concentration increases and induces the presence of bimerons. We found that even at zero temperature, the
number of skyrmions decreases rapidly as $\rho$ increases. To see this
behavior we have also plotted the skyrmion number as a function of
$\rho$ (see Fig.\ref{fig:qm/l4}). For a fixed very low temperature ($T=0.01J/k_{B}$), one can easily
observe that the skyrmion number decreases monotonically down to its
minimum value ($<N>_{sky}=0$) achieved in a critical $\rho=\rho_{c} \approx
0.35$, where skyrmions are not present in the system anymore.

\begin{figure}[hbt]
\includegraphics[height=38.0mm]{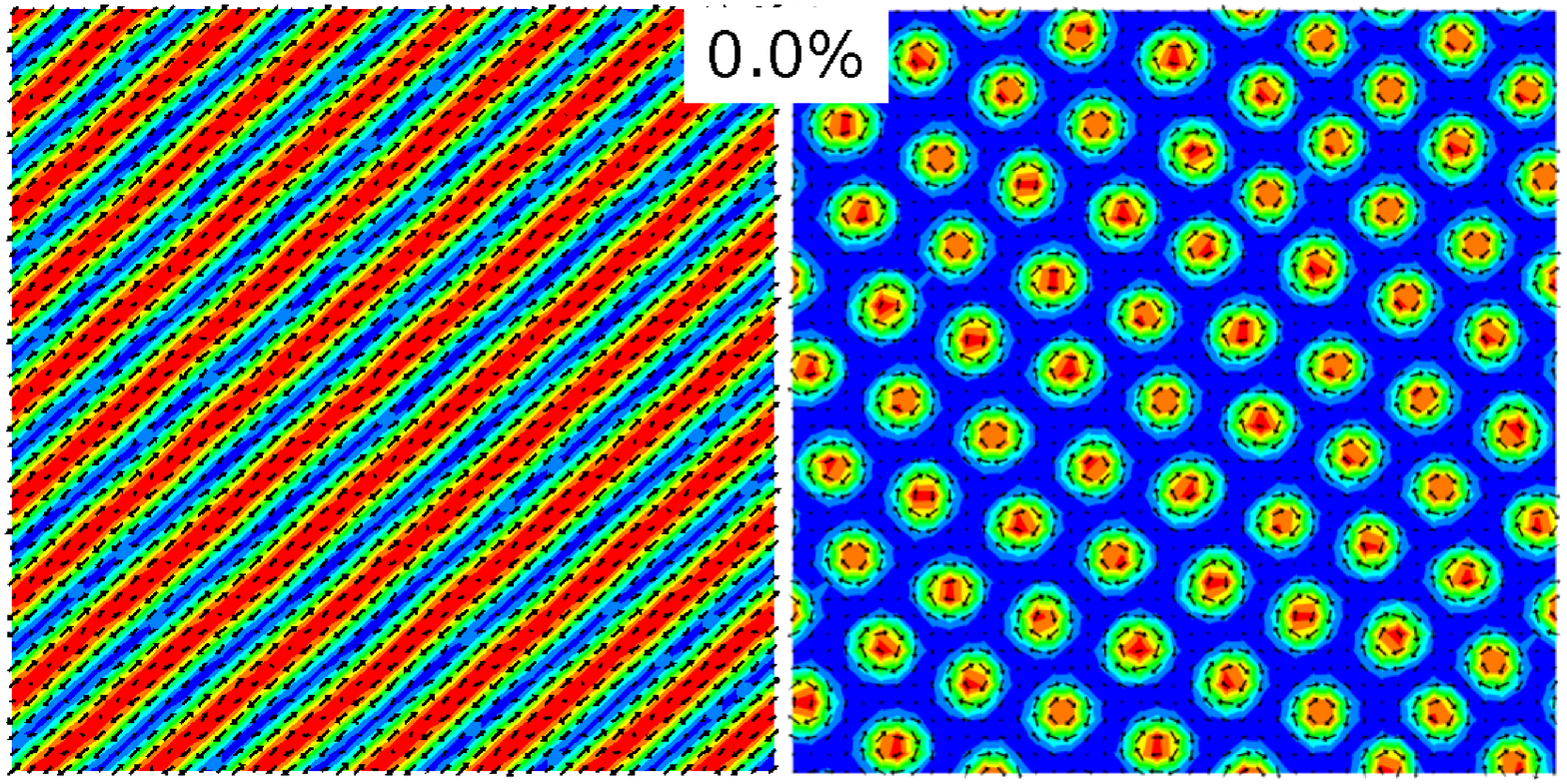}
\includegraphics[height=38.0mm]{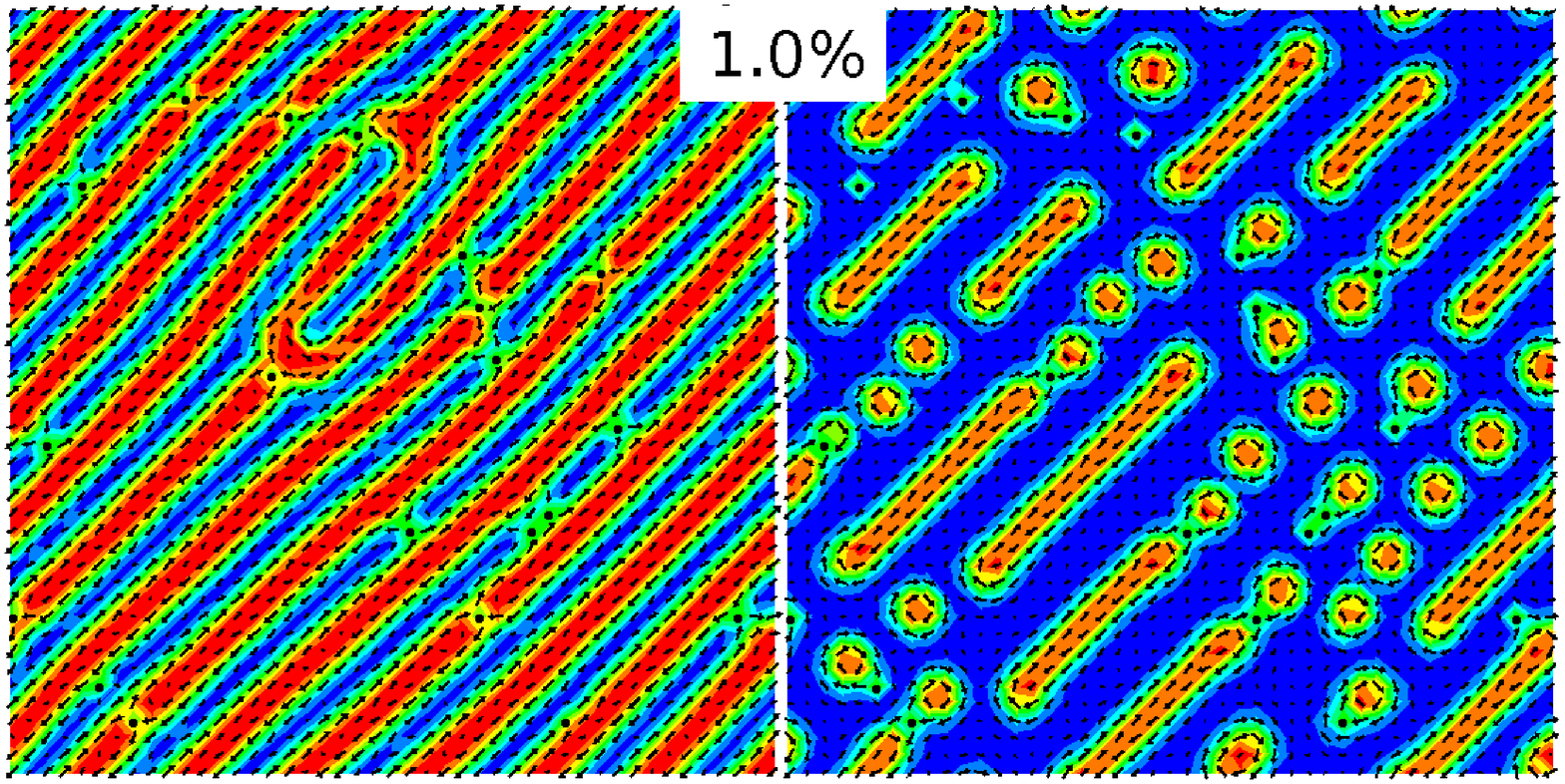}
\includegraphics[height=38.0mm]{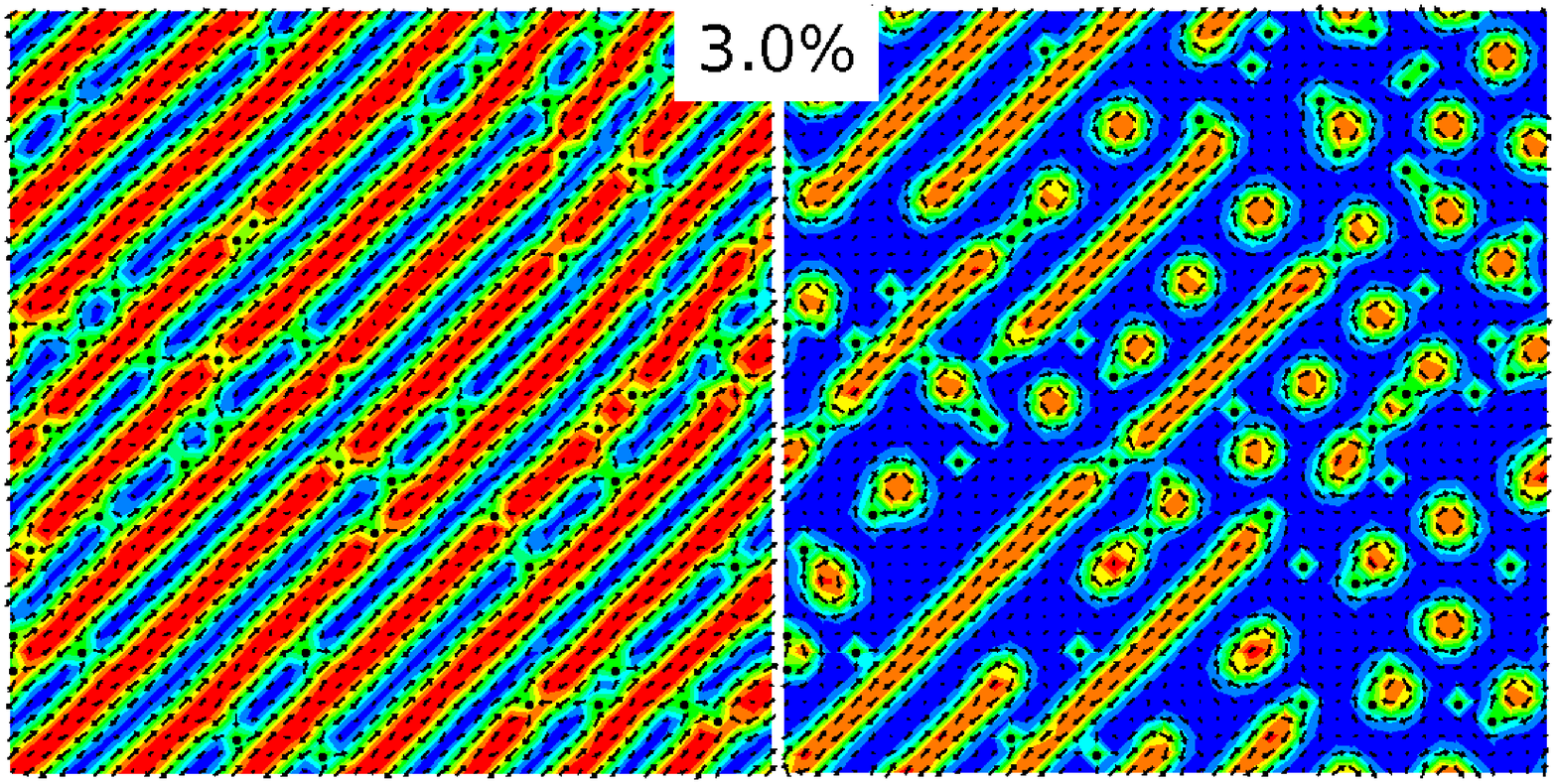}
\includegraphics[height=38.0mm]{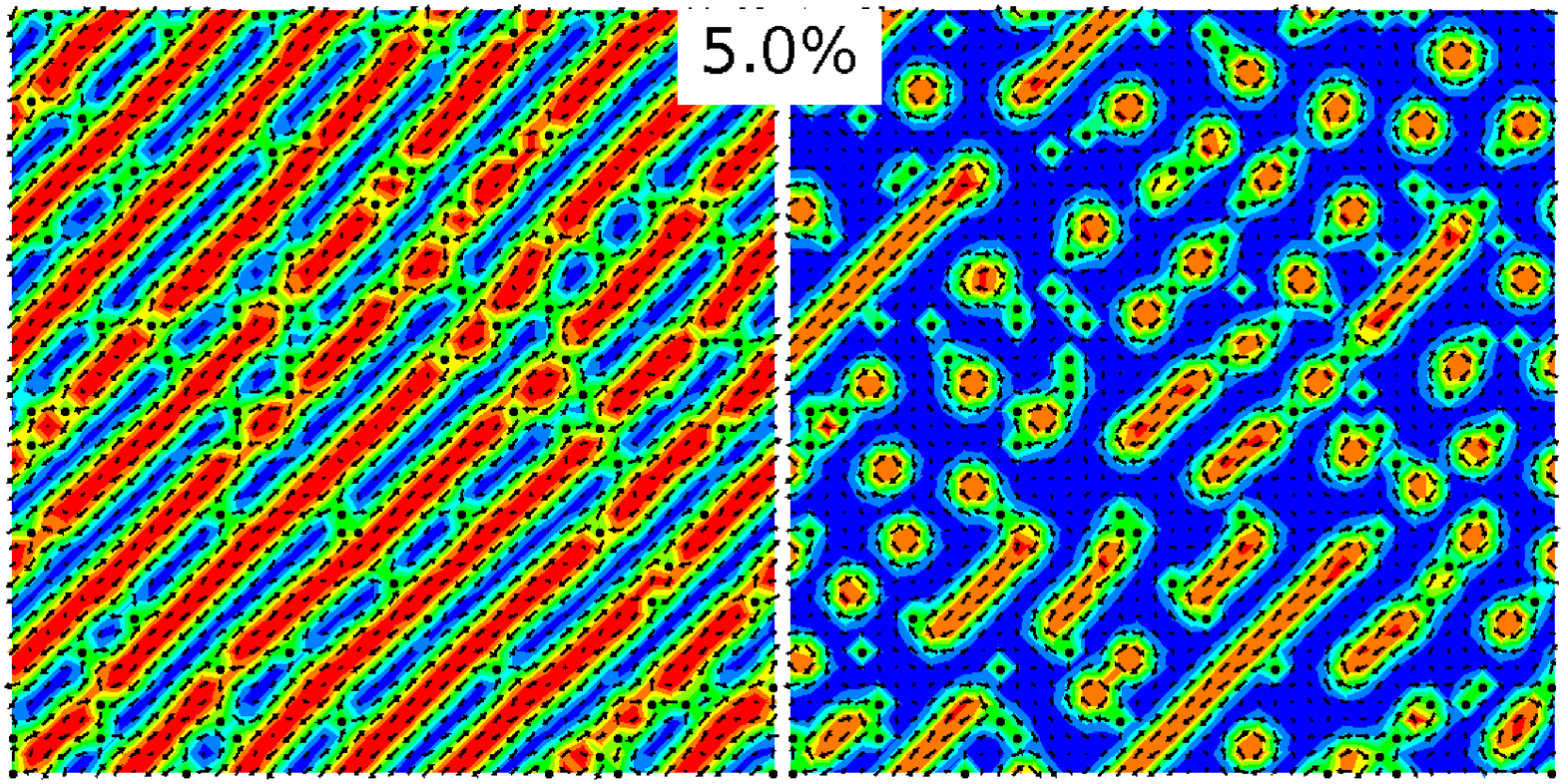}
\includegraphics[height=38.0mm]{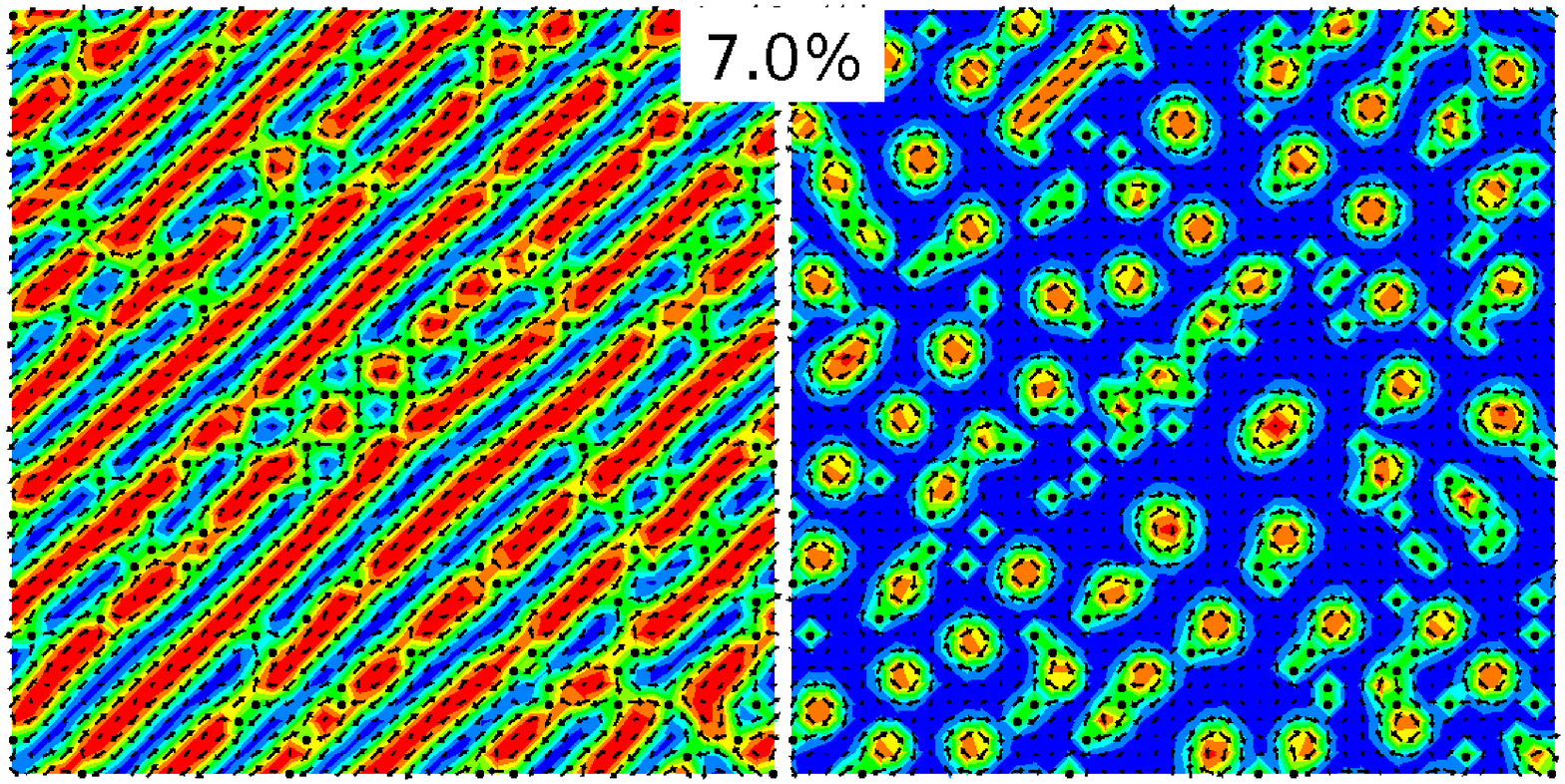}
\caption{\label{fig:qm/l8} The spiral and 
skyrmion states in a linear sequence with 
different concentrations of impurities.}
\end{figure}

\begin{figure}[hbt]
\begin{center}
\includegraphics[height=80.0mm,width=80.0mm]{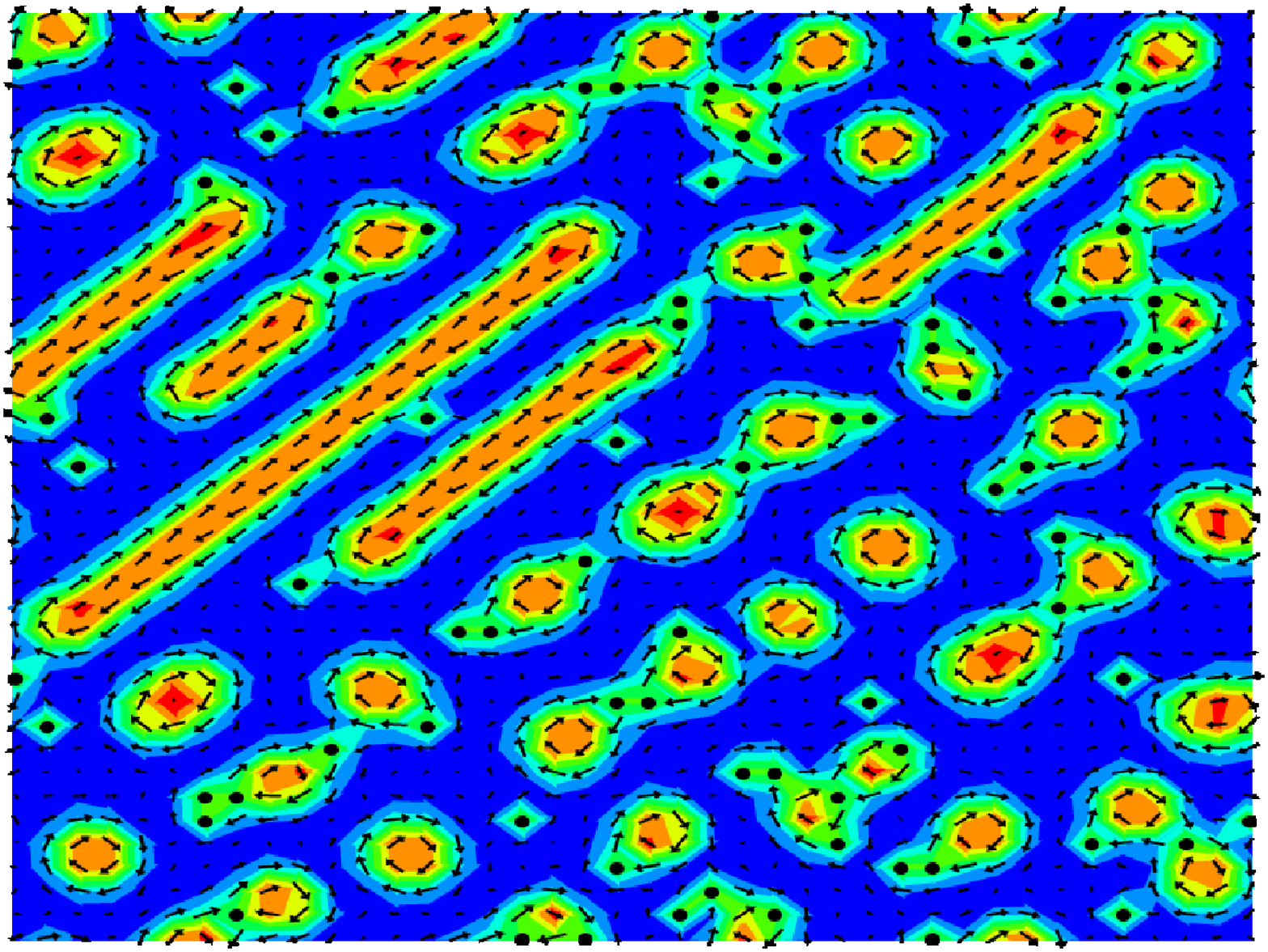}
\caption{\label{fig:qm/l6} Typical configuration of a $2D$ chiral magnet containing $1\%$ of impurities.}
\end{center}
\end{figure}

This critical value $\rho_{c}$ practically does not depend on the
lattice size. One of the reasons for the reduction of the skyrmion number is
that bimerons also appear in the skyrmion crystal and some of them occupy large
space regions which could contain $4$ or more skyrmions (there are also smaller
bimerons, occupying a space of $2$ or $3$ skyrmions). We also notice that the
six fold order is lost even for low concentration of vacancies. In fact, it is easy to see in
Fig.\ref{fig:qm/l4} that the skyrmion number reduces to almost
$1/3$ for $\rho=0.05$ and to almost $1/2$ for $\rho=0.1$ (at $T=0.01J/k_{B}$). Such large
lost of skyrmions can not sustain the perfect six folder order along
all the system. Our calculations indicate that a small concentration of
impurities (less than one percent) is already sufficient for destroying the six fold order,
deforming both the configuration of a large number of skyrmions and the skyrmion lattice
itself. In Fig.\ref{fig:qm/l8} we show the helical and skyrmion states in a sequence of impurities concentration.
Although isolated skyrmions are attracted to the spin
vacancies \cite{Subbaraman98,Mol03,Pereira03} in isotropic $2D$ Heisenberg magnets, which allow an impurity
to be at a skyrmion center, skyrmions in a skyrmion crystal (like bimerons in the helical state)
are not, in general, centered at vacancies. It is very common to see a sequence of defects
containing a compact meron, an impurity and a skyrmion along a diagonal line in the $xy$-plane (see Fig.\ref{fig:qm/l6} for the case of one percent of impurity at $T=0.01k_{B}/J$).  Therefore, in contrast to isolated skyrmions, energy minimization does not favor these objects to be pinned to spin vacancies when the skyrmion crystal is taken into
account. There are situations (more energetically favorable) and in this
case, the energy minimization must involve the whole crystal. We
notice that the skyrmion lattice is formed in such a way that the spin
vacancies become, in general, localized between two skyrmions or between a skyrmion and a bimeron around
the system lattice. In a perfect skyrmion crystal, the spins tend to
become out-of-plane in the regions between two skyrmions but when
impurities are present in these regions, they tend to put spins pointing
along the plane directions.

\begin{figure}[hbt]
\begin{center}
\subfigure[]{
\includegraphics[height=80.0mm]{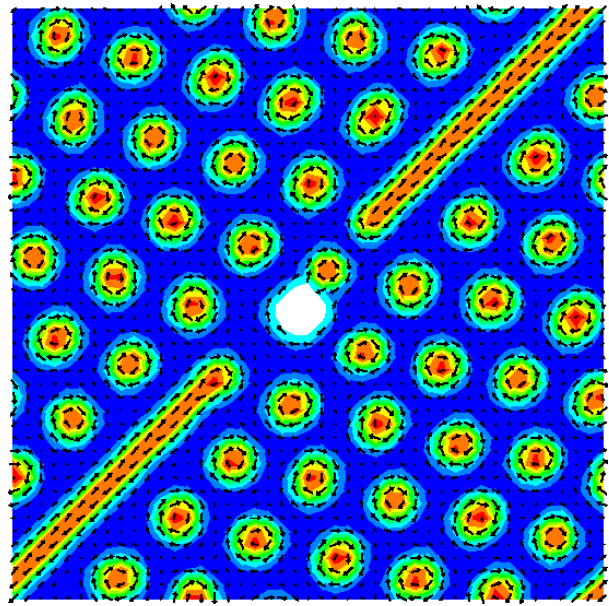}}
\subfigure[]{
\includegraphics[height=80.0mm]{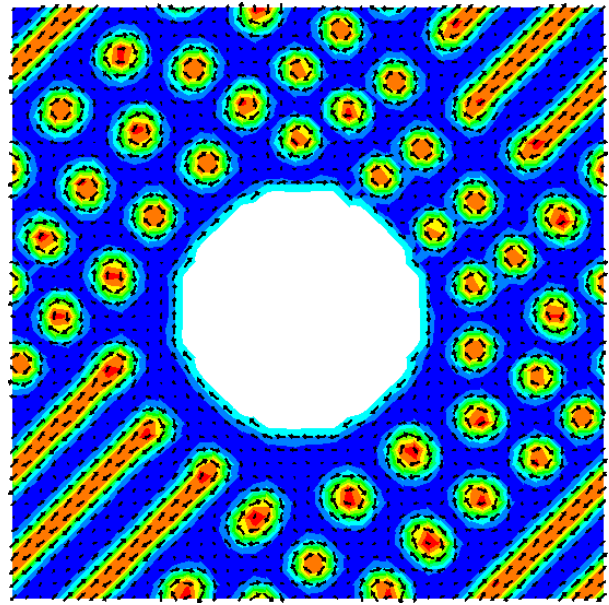}}
\caption{\label{fig:qm/l7}  (a): Typical configurations of a $2D$ chiral magnet containing a hole in its center (some adjacent spins removed forming an almost circular defect with radius $\kappa$). (a): $\kappa=2a$. (b): $\kappa=9a$.}
\end{center}
\end{figure}

We have also investigated the influence of defects containing a number of adjacent impurities removed from the system. These are hexagonal-shape defects introduced in the sample plane turning it out a non-simply connected manifold. Such larger defects may be intentionally introduced into the system. We have studied the cases in which the hole has radius $\kappa$ raging from one lattice constant $a$ (four spins removed) to $9a$ ($256$ spins removed). Figure \ref{fig:qm/l7} shows the situations for $\kappa=2a$ and $\kappa=9a$. In general, for small holes ($\kappa \leq 6a$), the six fold order is not destroyed for regions not so near the defect. Along the diagonal line passing for the hole center, it is common to surge bimerons. Around the hole border, the spins tend to become in the plane developing a vortex-like configuration. On the other hand, for relatively large holes ($\kappa \geq 7a$), bimerons start to appear even along diagonal lines very apart, evidencing the nonlocal influence of such defects. The 
situation is similar to that obtained with impurities and external perpendicular field (see Fig. \ref{fig:qm/l5} (b) and related text) where skyrmion six fold order and elongated stripes (bimerons) coexist. The main novelty here is that larger defects tend to preserve skyrmions in regions around the hole.

\section{Conclusions and Prospects}

Here, we have performed Monte Carlo simulations for diluted $2D$ chiral magnets; the helical state and the skyrmion
lattice were investigated. We have found that the randomly insertion of nonmagnetic impurities
into the system induces the occurrence of bimerons in the helical state (ground state) by breaking some stripe-domain structures. These compact merons also surge in the skyrmion crystal phase, decreasing the number of skyrmions. Impurities distort both the skyrmion configuration and the skyrmion lattice. Therefore, spin vacancies may be faced as an intrinsic mechanism inducing phase transition from skyrmion lattice to spiral state where elongated spin stripes show up abundantly throughout the system. In this sense, they appear to work similar to thermal effects in pure materials (see Ambrose and Stamps \cite{Ambrose12}). It is also noteworthy to mention that the same overall effect is obtained by the application of suitable magnetic field, as suggested by Ezawa \cite{Ezawa11}, and eventually, as observed in experiments with magnetic field cooling by Milde {\it et al} \cite{Science340}. Here, we have shown that the same happens as spin vacancies are continuously added to the system. As impurity concentration 
reaches a critical value $\rho_{c}\approx 0.35$ (regardless the lattice size), vestiges of skyrmion disappear completely and the system cannot support them anymore. We have also investigated how a non-simply connected manifold could affect the skyrmion crystal by considering adjacent impurities forming a large circular-type hole. All these types of defects like spin vacancies may affect several properties of the system; for instance, since a net chiral charge is present for a relatively large range of impurity density ($\rho < 0.35$) then, at a fixed temperature, the Hall conductivity may present interesting different consequences for different concentrations of impurities. As a final remark, we would like to mention that spin vacancies may offer additional advantages for probing topological objects, including skyrmions, merons, and bimerons, in chiral magnets. Actually, by combining magnetic microscopy setups with magneto-mechanical apparatus, like those used to investigate Barkhausen effects \cite{
Science2013-exp}, not only their direct detection as well as their precise location could be determined with very good accuracy.


.
\begin{acknowledgments}
The authors thank CNPq and FAPEMIG for
financial support.
\end{acknowledgments}

\end{document}